\documentclass[pre,a4paper,oneside,nofootinbib]{revtex4}

\usepackage[english]{babel}

\usepackage[letterpaper,top=2cm,bottom=2cm,left=3cm,right=3cm,marginparwidth=1.75cm]{geometry}

\usepackage{amsmath}
\usepackage{graphicx}
\usepackage[colorlinks=true, allcolors=blue]{hyperref}

\usepackage[normalem]{ulem} 

\begin{document}

\title{One-Dimensional Urban Scaling}

\author{Fabiano L.\ Ribeiro [0000-0002-2719-6061]}
\email{fribeiro@ufla.br}
\affiliation{Department of Physics (DFI), Federal University of Lavras (UFLA), Lavras MG, Brazil}

\author{Renan L. Fagundes [0009-0006-1677-4168]}
\email{r.fagundes@ioer.de}
\affiliation{Leibniz Institute of Ecological Urban and Regional Development (IOER), Dresden, Germany}

\author{Diego Rybski [0000-0001-6125-7705]}
\email{ca-dr@rybski.de}
\affiliation{Leibniz Institute of Ecological Urban and Regional Development (IOER), Dresden, Germany}
\affiliation{Complexity Science Hub Vienna, Josefst\"adterstrasse 39, A-1090 Vienna, Austria}


\begin{abstract}
In this paper, we apply recent findings from urban scaling theory to evaluate how it could be applied to a one-dimensional archetypal city. Our focus is on how the simplicity of a one-dimensional model can provide intuitive insights that might be obscured in more complex, multidimensional models.
With this instructive model, it is possible to visualize that if the population's accessibility decreases more slowly than the increase in distance, an agglomeration effect occurs. This leads to increasing returns to scale when considering interactions between people or economies of scale in the case of amenities that serve the city.
These findings highlight a key characteristic of complex systems: interacting parts that together exhibit properties that are not evident when considered individually.
The results presented here are derived from both numerical simulations and theoretical analysis.
Although one-dimensionality is used here as a simplification, many natural (e.g., Sułoszowa in Poland) and artificial 
 cities (e.g., The Line in Saudi Arabia)  worldwide are shaped this way. 
\end{abstract}

\maketitle

\section{Introduction}

Urban scaling theory describes the relationships between the size of cities and various urban metrics \cite{bettencourt2007growth}. 
The power-law scaling $Y \sim N^{\beta}$ states that certain urban variable $Y$ is related to the city population size $N$ raised to a \emph{scaling exponent} $\beta$. There are three main categories of urban scaling variables. The first is the \emph{superlinear scaling}, $\beta >1$, usually associated with socioeconomic variables. In this case, more populous cities exhibit larger socioeconomic metrics per capita, which in economics is known by \emph{increasing return to scaling}. This can be observed for GDP, income, and innovation indicators. The second category is the \emph{sublinear scaling}, $\beta <1$, related to the infrastructure variables. In this case, cities with larger populations need less infrastructure per capita, revealing a scaling economy. 
For example, this applies to built-up areas, the number of petrol stations, or other types of amenities \cite{Kuhnert2006}. 
Lastly, we have the \emph{linear scaling}, $\beta = 1$, which refers to variables of individual needs. These variables represent situations in which the size of the city does not interfere with the per-capita values, e.g., water consumption or the number of employees. In summary, the scaling law provides insight into how different variables relate to city population size, and a recent review of this topic can be found in \cite{Ribeiro_chapter2024,Compendium2025}.

Many empirical \cite{Andris2014,Schlapfer2014,Samaniego2020} and theoretical \cite{Bettencourt2013,Molinero2021,ribeirocity2017} studies suggest that the increasing returns in socio-economic variables result from the social interaction between people, which increases faster than the city’s population size.
Some works also argue that the super-linear growth of social interactions can be quantitatively explained by assuming that pairwise interactions between individuals decay with distance, following some form of \textit{gravity model} 
\cite{ribeirocity2017,yakubo2014,Arbesman2009}. 
Gravity models, particularly in their simplest form explored in the present work, have a long history in urban modeling, dating back over 170 years \cite{Philbrick1973}.
These models have been fundamentally shaped and stabilized in the context of urban phenomena since the establishment of Tobler’s law \cite{Tobler1970}: ``\textit{Everything is related to everything else, but near things are more related than distant things.}''
Examples of their application in urban phenomena include not only urban scaling \cite{Ribeiro2023,Barthelemy2019} but also commuter patterns \cite{Spadon2019}, 
social media and mobile phone contacts
\cite{Goldenberg2009,Schlapfer2021,Herrera-Yague2015}, migratory flows \cite{Curiel2018}, and many others.

We are interested in investigating specific properties of urban scaling in a one-dimensional scenario.
We assess the theoretical simplicity of a one-dimensional archetypal city and conduct several analyses.  As will be shown, the mathematical complexity encountered in higher dimensions becomes more straightforward in a one-dimensional context without losing any fundamental aspects.  
One-dimensional analysis proves particularly useful by presenting important concepts related to urban scaling in a simple and accessible manner.
These concepts might be difficult to perceive or investigate in higher-dimensional contexts.

\begin{figure}
\includegraphics[width=1\linewidth]{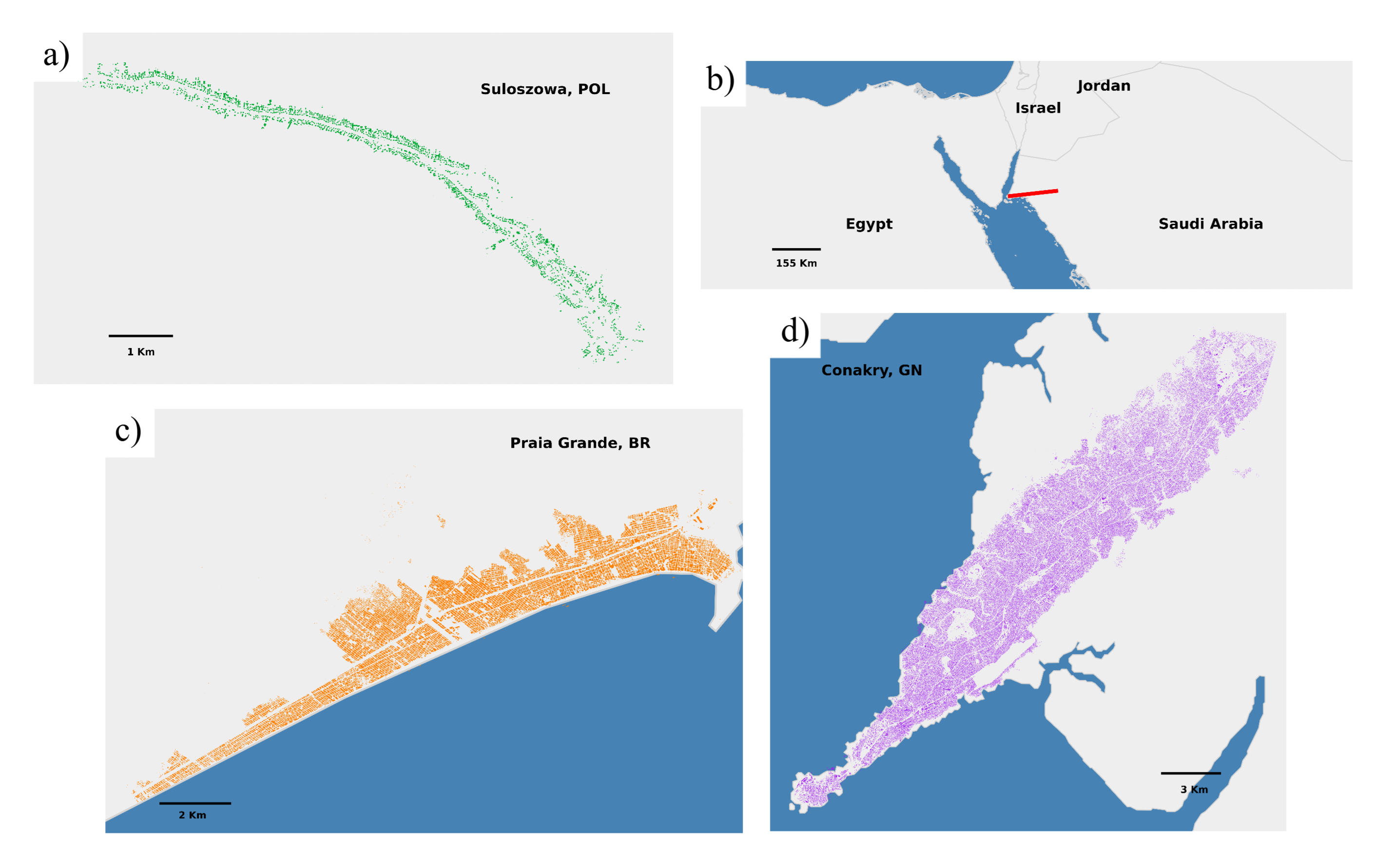}
\caption{\label{fig:the_line}
Examples of (approximately) one-dimensional cities. a) Sułoszowa (green), in Poland. b) The Line (in red), in Saudi Arabia. c) Praia Grande (orange), on Brazilian cost. d) Conakry (purple), in Guinea. 
}
\end{figure}

An example of a one-dimensional city is \emph{The Line}, a city that is presently being built in Saudi Arabia; see Fig.~\ref{fig:the_line}(b).
It is planned to extend over a length of 170 kilometers, measuring 200 meters in width and reaching a height of 500 meters. 
It corresponds to a wall consisting of several \emph{ World~Trade~Centers} (which was 541,3 meters high), side by side, extending from San~Diego to Los~Angeles, California.
It is planned to accommodate 9 million inhabitants within a total area of 34\,km$^2$, 
resulting in a projected surface density of 265,000 people per km$^2$ and a linear density of 53~thousand people per kilometer \cite{Prieto-Curiel-circle2023}.
Just to compare, Manhattan's population density is approximately 27,826 people per 
km$^2$, which makes it one of the most densely populated areas in the United States and the world, but it is ten times less dense than The Line. 
The Line, as part of \emph{Neom}, is an example of a \emph{planned city}, and some recent works have discussed the challenges of this kind of urban settlement \cite{Prieto-Curiel-circle2023,Batty2022,Rivas2023}. 
However, there are also many natural-emergent   
one-dimensional cities, such as \emph{Suloszowa}, in Poland, which is formed by a single street that spans approximately 8 kilometers and houses 5,800 inhabitants. 
Or \emph{Praia Grande} at the Brazilian coast, as shown in Fig.~\ref{fig:the_line}(c).
In contrast to The Line, these natural $\sim 1$D  
cities are strongly shaped by the geographies in which they developed and grew.

\section{Social Interaction in a 1D city}

We examine a theoretical city with a population of size $N$ homogeneously distributed along a line of length $L$. Let’s assume that the interaction between individuals in this city follows a gravity model.

\subsubsection*{Computational Model}

To simulate and examine interactions among people in this archetypal city, we propose the following model, which allows us to better understand how the number of contacts scales with the city's population size.
For this purpose, consider the individual $i$ who lives in the position $x_i$; the probability that he/she visits a part of the city which is at a distance $r = |x-x_i|$ from him/her is given by 
\begin{equation}\label{eq_pi}
   p_i(x) \propto  \frac{1}{|x-x_i|^{\gamma}}, \, \, \textrm{  for $|x-x_i|>r_0$} \, .
\end{equation}
Here, $r_0$ is a small distance containing one single person,
and $\gamma$ is the \emph{decay exponent}, which governs the interaction range.
A similar gravity model, analyzed in the context of population growth within a one-dimensional scenario, is discussed in \cite{ribeiro1d}.

\begin{figure}
\centering
a) \includegraphics[width=0.6\linewidth]{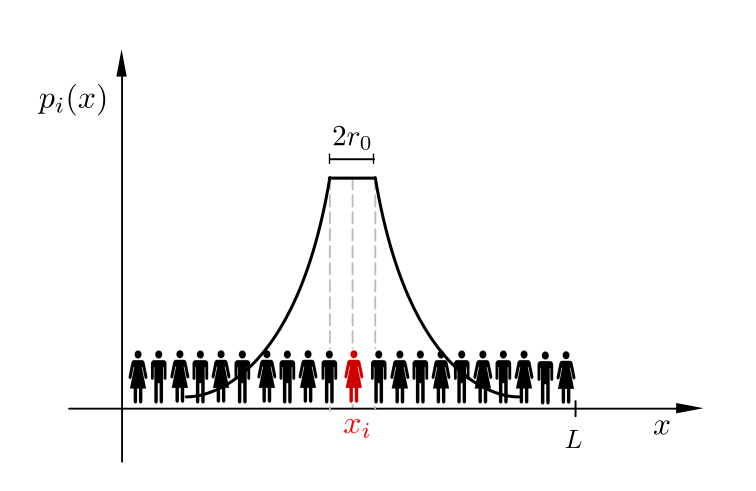} \\
b) \includegraphics[width=0.3\linewidth]{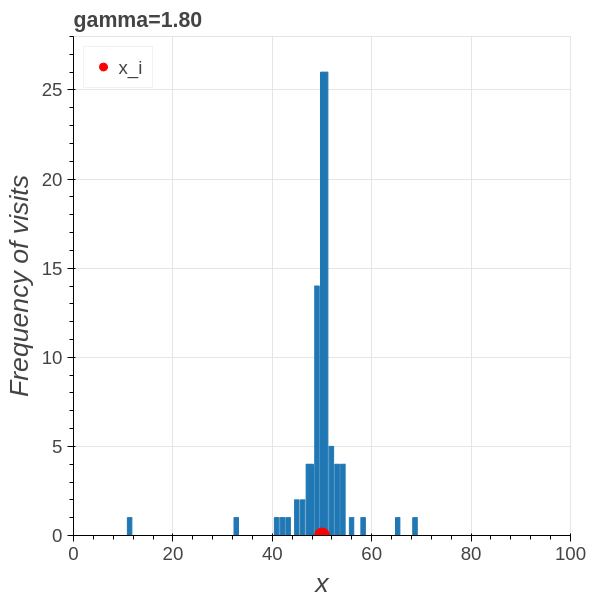}
c) \includegraphics[width=0.3\linewidth]{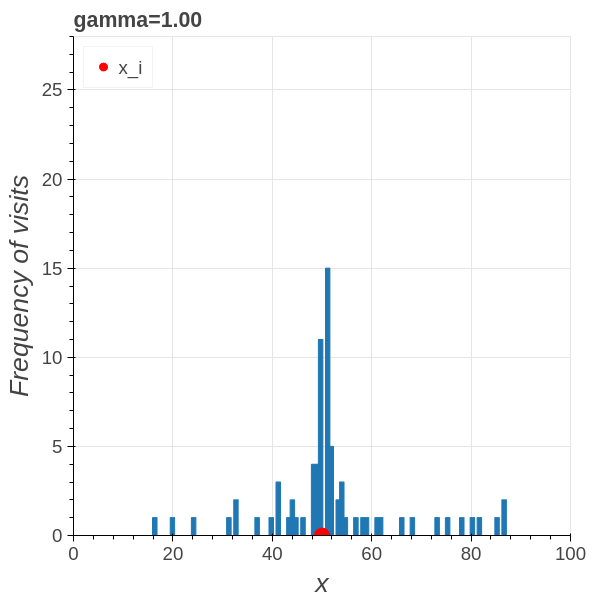} 
d) \includegraphics[width=0.3\linewidth]{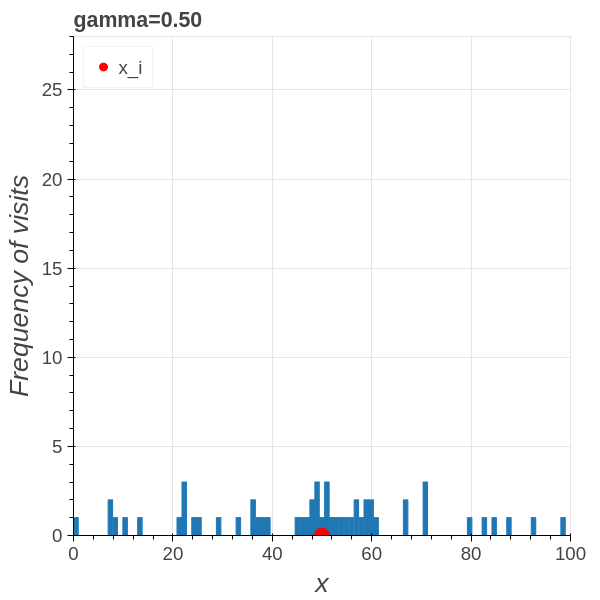}
\caption{\label{fig:histo_visits}
Illustration of the probability that the $i$-th individual, residing at position $x_i$, will visit location $x$ in a one-dimensional city of size $L$.
When this individual visits a specific place, he/she becomes connected (establishes a link) with all the individuals that live there.
The panel a) illustrates the behavior of this probability, say $p_i(x)$, according to Eq.~\ref{eq_pi}. 
Numerical simulation examples, using $x_i = 50$ and $L = 100$, are shown in panels (b), (c), and (d). These panels illustrate the frequency of visits by the individual $i$ to position $x$ after 100 computational steps, highlighting the variation in movement patterns based on the decay exponent $\gamma$.
At each location visited (regardless of frequency), the individual establishes contact with the residents of that location.
For a sufficiently large $\gamma$ (panel b, $\gamma = 1.8$), the individual is restricted to visiting closest neighbors' positions, representing a short-range interaction regime. Conversely, for a sufficiently small $\gamma$ (panel d, $\gamma = 0.5$), the individual visits locations across the entire city, representing a long-range interaction regime. Panel c ($\gamma = 1$) shows an intermediate scenario, marking the transition between the two regimes.
}
\end{figure}

To simplify the analyses, suppose there is one person per unit of length (as depicted in Fig.~\ref{fig:histo_visits}-a).
In each (computational) time step, $i$ visits a location based on the probability $p_i(x)$ (a random walk with a return to the origin $x_i$ at each time step). 
The individual $i$ establishes a connection (contact) with all the people that reside in the location he/she visits. After a given number of computational steps, it is possible to compute how many people $i$ visits/meets.
Then, after a certain number of steps, $i$ will have $k_i$ contacts; 
that is, $k_i$ is the network degree of $i$.
Figure~\ref{fig:histo_visits}-b)-c)-d) illustrates the movement (simulation) of a single individual through the one-dimensional city.

\begin{figure}
\centering
a) 
\includegraphics[width=0.45\linewidth]{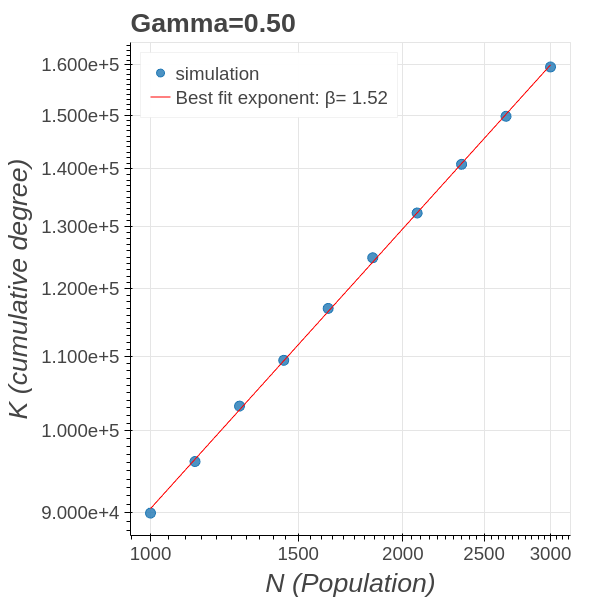}
b)
\includegraphics[width=0.45\linewidth]{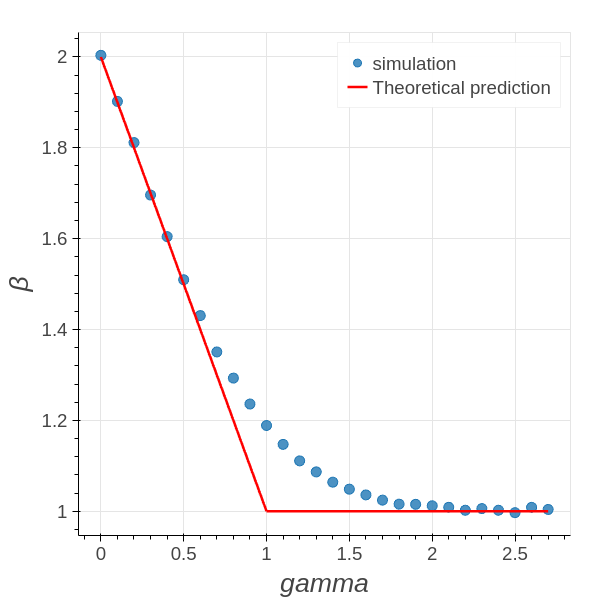} \\
\caption{\label{fig:superlinear-scaling} 
Computational simulation and comparison with the theoretical prediction of the proposed model (described in the main text). Panel a) shows how the cumulative degree $K$ scales with the population size in the computational simulation of the model (blue dots) for a specific value of the decay exponent ($\gamma = 0.5$). The fit to the dots by the power law $K \sim N^{\beta}$ (red line) allows us to get the scaling exponent $\beta$ that emerges from the simulation. 
Panel b) presents the value of the numerical scaling exponent for various values of $\gamma$ and compares it with the theoretical result (from Eq.~(\ref{eq_resultado_super})). 
For $\gamma > 1$, which represents a short-range interaction regime, the scaling is linear ($\beta = 1$), whereas for $\gamma < 1$, representing a long-range interaction regime, superlinear scaling emerges. 
}
\end{figure}

We aim to investigate how the \emph{average degree}, denoted as
 
\begin{equation}
    \langle k \rangle = \frac{1}{N} \sum_{i=1}^N k_i  \, , 
\end{equation}
and the \emph{cumulative degree} in the city, say $K = N  \langle k \rangle$, behave according to the interaction range imposed by $\gamma$ and how they scale with the population size.
Fig.~\ref{fig:superlinear-scaling} presents simulation results stemming from this model, indicating that the total number of contacts grows superlinearly with the city population size ($\beta>1$) when $\gamma$ is sufficiently small, corresponding to a scenario where the interaction range is large enough (as represented in Fig.~\ref{fig:histo_visits}-d)).
In addition, the simulation shows that linear scaling ($\beta =1$) occurs when $\gamma$ is sufficiently large, corresponding to a scenario where the interaction range is very small (as represented in Fig.~\ref{fig:histo_visits}-c)).

\subsubsection*{Theoretical Analysis}

Let us solve the model analytically to understand what is happening. To do this, consider $dk_i(x)$ as the number of contacts of $i$ that live between $x$ and $x+dx$, which can be computed using

\begin{equation}
 dk_i(x) = p_i(x) dN(x)    
 \, ,
\end{equation}
where $dN(x)$ is the number of people that lives between $x$ and $x+dx$. If there is one person per unit of length (homogeneous distribution of people in 1D space), we can write that $dN(x) = dx$. Consequently $dk_i(x) = p_i(x) dx$.
Then, the average degree of $i$ can be computed as 

\begin{equation}
    \langle k_i \rangle  = \int_{r_0}^{L} dk_i(x) = \int_{r_0}^{L}p_i(x) dx
    \, .
\end{equation}
Using the definition Eq.~(\ref{eq_pi}) and considering, without loss of generality, that $x_i = 0$ and $r_0 =1$ (to exclude self interaction) yields

\begin{equation}
    \langle k_i \rangle  \sim  \frac{1}{(1 - \gamma)} \left[ N^{1-\gamma} -1  \right]
    \, ,
    \end{equation}
where we also consider that $N = L$ (one person per unit of length).  
Consequently, the cumulative degree $K = N \langle k_i \rangle$ scales with the population size as 
\begin{equation}\label{eq_K}
   K \sim  \frac{1}{(1 - \gamma)} \left[ N^{2-\gamma} -N  \right]
   \, .
\end{equation}
Note that if $\gamma>1$, the second term on the left of Eq.~(\ref{eq_K}) dominates when $N$ is sufficiently large; then $K \sim N$: linear scaling. 
In this situation ($\gamma>1$), the interaction decay function decreases so quickly with the distance that we refer to this situation as the \emph{short-range interaction regime} \cite{Campa2009,Campa2014,ribeirocsf2024}. 
The other case, $\gamma<1$, the first term on the left of Eq.~(\ref{eq_K}) dominates for $N$ sufficiently large; then $K\sim N^{\beta}$, with $\beta = 2-\gamma$, that is superlinear scaling.
In this situation ($\gamma<1$), the interaction decay function decreases so slowly with the distance that we refer to this situation as the \emph{long-range interaction regime}.
We can sum up this result (for $N\rightarrow\infty$) as

\begin{equation}\label{eq_resultado_super}
 K \sim  \left\{ \begin{array}{ll}
N^{\beta} & \textrm{if  $\gamma<1$ (long-range regime) with  $\beta =  2- \gamma$}\\
 &  \\
N & \textrm{if  $\gamma>1$ (short-range regime)  }
\end{array} \right. \, .
\end{equation}
This theoretical prediction aligns well with the numerical simulation shown in Fig.~\ref{fig:superlinear-scaling}, demonstrating a fair agreement between the two, except near $\gamma=1$ (the critical point) due to finite size effects.
Eq.~(\ref{eq_resultado_super}) is the particular case ($D=1$) of the result $\beta = 2- \gamma/D$ presented and discussed in \cite{ribeirocity2017}, where $D$ is the fractal/euclidean dimension of the city.

The result in Eq.~(\ref{eq_resultado_super}) allows the following interpretation. The case $\gamma = 1$ distinguishes between two regimes: the short-range interaction regime ($\gamma>1$) and the long-range interaction regime ($\gamma<1$).
In the long-range interaction regime, the region of the city accessible to each individual scales with the size of the city. In this context, we could say that people interact effectively with the city as a whole, as shown in Fig.~\ref{fig:superlinear-scaling}-d). This is the case where increasing returns to scale occur -- in the sense that when the city's population increases, the number of contacts within the city increases by more than a proportional amount.
The other case, in the short-range interaction regime, people limit their interactions to nearby neighbors (as shown in Fig.~\ref{fig:superlinear-scaling}-c). In this case, since individuals are restricted to interacting with their immediate neighbors, the number of contacts for each person saturates and does not increase with city size. This situation leads to the number of contacts scaling linearly ($\beta = 1$) with population size. This suggests that in this short-range interaction regime, the city essentially functions as a collection of isolated groups, preventing the emergence of increasing returns to scale.

\section{Infrastructure: optimal amenities number}

\begin{figure}
\centering
a) 
\includegraphics[width=0.46\linewidth]{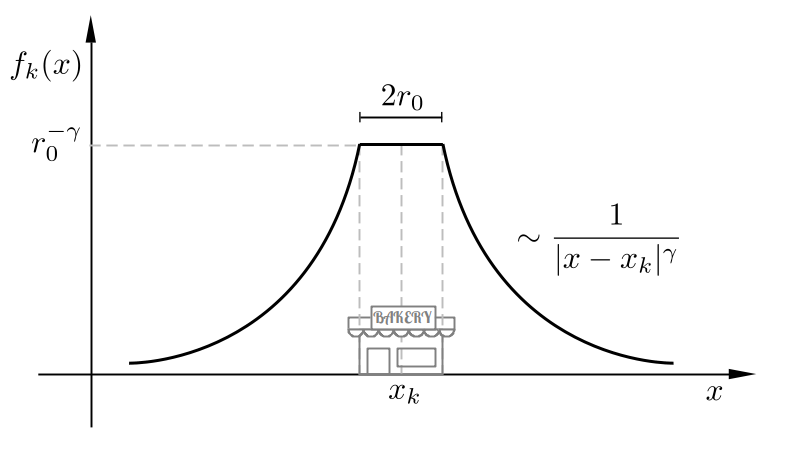}
b)
\includegraphics[width=0.46\linewidth]{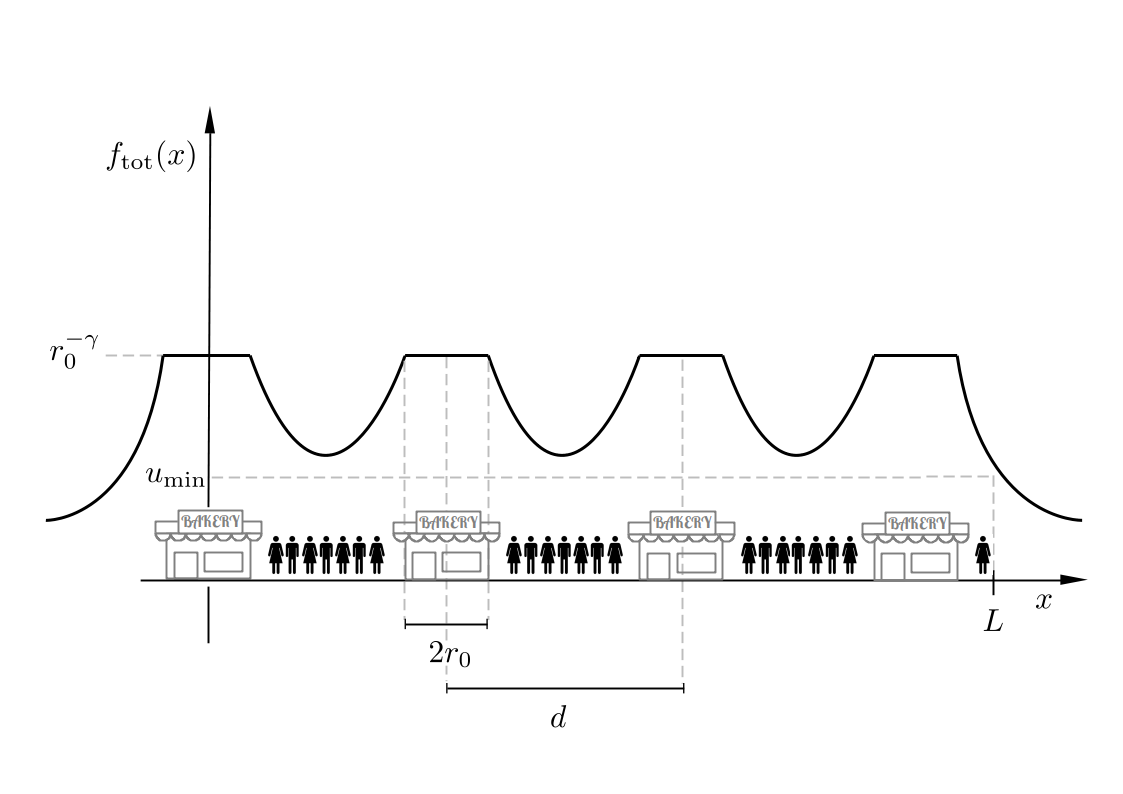}
\caption{\label{fig:fk} 
Illustration of the amenities (e.g., bakeries) in the one-dimensional city model. 
a) The influence $f_k(x)$ of the $k$-th amenity (located at $x_k$) on the position $x$. b) Influence aggregation of all amenities in the archetypal one-dimensional city. 
If $u_{\rm{min}}$ is the minimum intensity of influence required for a part of the city to function effectively, then for the entire city to operate well, it is necessary that $f_{\rm{tot}}(x) \ge u_{\rm{min}}$ for all $x \in L$. In an ideal scenario, where there is neither excess nor scarcity of service, we have
$\min_x \{f_{\rm{tot}}(x) \} \approx u_{\rm{min}}$.
}
\end{figure}

Social interactions only represent one side of the coin, and infrastructure scales differently.
Thus, next, we want to use similar ideas to describe the infrastructure in the city.
For this purpose, we continue to use the archetypic, one-dimensional city, assuming a homogeneous space distribution of a single type of amenity (e.g., bakery). Furthermore, each amenity has a range of interactions (or range of services) expressed by a decay function of the following kind
\begin{equation}\label{eq:model_fk}
f_k (x) =  \left\{ \begin{array}{ll}
\frac{1}{|x-x_k|^{\gamma}} & \textrm{if $|x_k - x| > r_0$} \\ \\
r_0^{-\gamma} & \textrm{otherwise } \, ,
\end{array} \right.
\end{equation}
where $f_k(x)$ represents the intensity of the influence of the $k$-th amenity (located at $x_k$) at the position~$x$, analogous to Eq.~(\ref{eq_pi}). 
Similar to the previous section, the decay exponent $\gamma$ governs the \emph{range of influence} of the amenity, and $r_0$ now represents the physical length of each amenity. 
``Influence'' can be understood as the likelihood or chance of a person living in the location $x$ to use, or be supplied by, a specific amenity in $x_k$ (e.g., purchasing a product: bread in the case of a bakery). Figure~\ref{fig:fk}-a) illustrates the behavior of the definition above.

When there are $P$ amenities of the same type (class) throughout the city, the total influence of all of them at a specific position $x$ will be $f_{\rm{tot}}(x) = \sum_{k=1}^P f_k(x)$. 
An illustration of how $f_{\rm{tot}}(x)$ looks in the special case of equally spaced amenities is presented in Fig.~\ref{fig:fk}-b).
In this case, a fixed distance $d$ separates every two neighboring amenities.

\subsubsection*{Computational Model}

For the city to function effectively, the amenities of a specific sector must adequately serve all citizens. 
This means that the aggregate influence $f_{\rm{tot}}(x)$ must be sufficiently large for all parts of the city. 
To account for this, let us consider $u_{\rm{min}}$ as the minimum value of the intensity of influence that is required by a part of the city to function well.
Therefore, for the entire city to work effectively, it is necessary that this condition holds true for all parts of the city, i.e., $f_{\rm{tot}}(x) \ge u_{\rm{min}}$ for all $x \in L$. 
In the ideal scenario, where there is neither excess nor scarcity of service, one has $\min_x \{f_{\rm{tot}}(x) \} \approx u_{\rm{min}}$.

Note that in Fig.~\ref{fig:fk}-b) an over-optimal situation is represented because $f_{\rm{tot}}(x) > u_{\rm{min}}$ for all $x$, indicating that all individuals are over-supplied.
To optimize resources, it is necessary to decrease the number of amenities, and consequently, the distance $d$ between them will increase (to maintain a homogeneous distribution). As a result, the aggregate influence curve $f_{\rm{tot}}(x)$ will also decrease vertically. After removing a sufficient number of amenities (akin to bankruptcy in a real situation), the system reaches the optimal state. Analogously, one could start with just one amenity and add new ones until reaching the optimal state. Given this rule, $\min_x \{f_{\rm{tot}}(x) \} \approx u_{\rm{min}}$ represents the equilibrium state that determines the ideal number of amenities the city needs to serve its residents in an optimized way.


This procedure can be performed numerically to observe, for instance, how the optimal number of amenities ($P$) scales with the city's linear size ($L$) and the population ($N$). 
We want to verify whether the urban scaling $P \sim N^{\beta}$ holds, and if so, determine the expected scaling exponent ($\beta$), given that the influence of the amenities is determined by $\gamma$, similar to the previous section.
Figure~\ref{fig:results}-a) shows numerical results and demonstrates that $P$ and $N$ follow a power-law relation (given by a straight line in a log-log plot), exhibiting sublinear behavior ($\beta < 1$) when $\gamma$ is sufficiently small (specifically when $\gamma < 1$, see Fig.~\ref{fig:results}-b)). 
However, the behavior becomes linear ($\beta = 1$) when $\gamma$ is sufficiently large ($\gamma > 1$).
Apparently, there is a transition at $\gamma=1$, similar to the previous section, which we want to understand in the following theoretical considerations.

\subsubsection*{Theoretical Analysis}

\begin{figure}
\centering
a)
\includegraphics[width=0.46\linewidth]{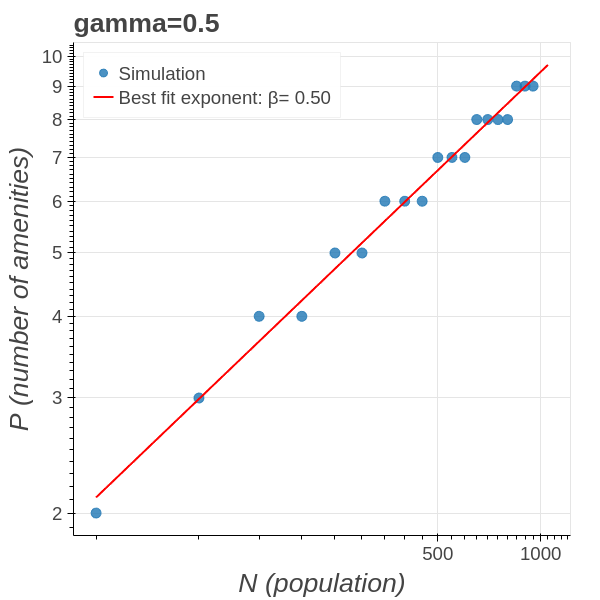}
b)
\includegraphics[width=0.46\linewidth]{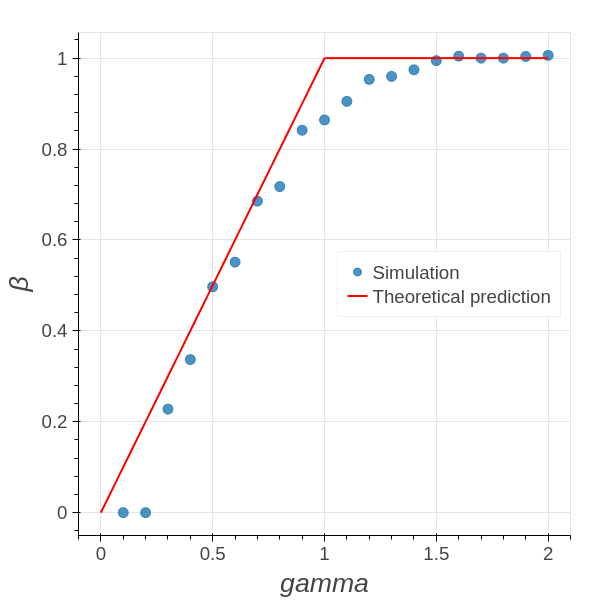}
\caption{\label{fig:results} 
Computational simulation and theoretical prediction of the scaling of amenities number in the one-dimensional city model.
a) Optimal number of amenities ($P$) as a function of the population size ($N \sim L$) for a specific value of the decay exponent $\gamma$ ($=0.5$); 
the dots represent numerical results, and the red straight line (best fit of the dots) is evidence of a power-law relation $P \sim N^{\beta}$. b) Scaling exponent $\beta$ as a function of $\gamma$, for both numeric (dots) and theoretical prediction (red lines). For $\gamma<1$ (long-range interaction),  $\beta \approx  \gamma $ and therefore the scaling is sub-linear; for $\gamma>1$ (short-range interaction),  $\beta = 1$ and therefore the scaling is linear.  
}
\end{figure}

To compare and test the numerical results, we consider a specific (one-dimension) case of the theoretical results presented in the references \cite{ribeirocity2017,Ribeiro2023}.
Let us also assume that the consumption of a product offered by the amenities of a given sector is given by $u(x)$. This could, for example, represent the number of breads necessary to serve the residents at position $x$ if the amenities considered are bakeries. It is reasonable to express $u(x) = \sum_{k=1}^P f_k(x)$, where $f_k(x)$ (influence of the $k$-th amenity at $x$) can also be interpreted as the number of products supplied by the $k$-th firm for the residents at $x$. 
Please note that the consumption expressed by $u(x)$ is analogous to the aggregated influence $f_{tot}$ used above in the context of the numerical model.
Then, the total consumption of this specific product in the city is $U = \sum_{x=0}^L u(x)$. Assuming that this product is an individual-need product implies that the average value of $u(x)$ is independent of the city size and, consequently, $U \sim N$ \cite{bettencourt2007growth}. 
Then it is possible to express $U$ as \cite{ribeirocity2017,Ribeiro2023}
\begin{equation}\label{eq:U}
U = \sum_{x=0}^L\sum_{k=1}^P f_k(x) = \sum_{k=1}^P  \left( \sum_{x=0}^L  f_k(x) \right) 
\, .  
\end{equation}
In the limit of continuous space, we can make the approximation \cite{ribeirocity2017,Ribeiro2023} 
\begin{equation}\label{eq:approx}
\sum_{x=0}^L  f_k(x) \approx 2 \int_{r_0}^L f_k(x -x_k) dx \sim \int_{r_0}^L x^{-\gamma}dx \sim \left(L^{1-\gamma} - r_0^{1-\gamma}\right)/(1-\gamma)
\, .
\end{equation}
The optimal number \( P \) is then the one that promotes \( U \sim N \). Thus, from Eqs.~(\ref{eq:U}) and~(\ref{eq:approx}),
and considering a homogeneous distribution of the population ($N\sim L$), one has $ N \sim U \sim P (N^{1-\gamma} - r_0^{1-\gamma})/(1-\gamma)$, and then we can conclude (in the limit of sufficiently large $N$) that
\begin{equation}\label{eq:PxN}
P \sim   \left\{ \begin{array}{ll}
N^{\beta}  & \textrm{if $\gamma < 1$ (long-range regime) with $\beta = \gamma$, } \\ \\
N & \textrm{if $\gamma > 1$ (short-range regime)} \, .
\end{array} \right.
\end{equation}
This result allows us to draw conclusions about the model and its consequences. 
First, the specific value \(\gamma = 1\) divides two distinct regimes, analogous to those observed in the context of the interaction between people described in the previous section.  When \(\gamma > 1\) (short-range interaction regime), 
the number of amenities scales linearly with the population size, \(P \sim N\). 
However, when \(\gamma < 1\)  (long-range interaction regime) the number of amenities scales \emph{sublinearly} with the population size, \(P \sim N^\beta\), with \(\beta = \gamma\).
Is the interactions decay slower than the gain of space along the linear extent, then the amenities' interactions add up.
This theoretical prediction aligns with the numerical simulation results presented in Fig.~\ref{fig:results}. In the general case, for a city with dimension \(D\), one can demonstrate that these ideas lead to \(P \sim N^{\gamma/D}\) \cite{ribeirocity2017}.

These results led us to interpret that economies of scale, in the sense of reducing per capita quantities with population growth, only occur in a regime of long-range interactions. In other words, when the components interact within a well-defined radius of action, which in practice means a situation of semi-isolated neighborhoods, economies of scale do not emerge. 
Economies of scale arise only when the city is effectively integrated, which is represented here by the long-range interaction regime.

Here, we use the example of bakeries and bread to illustrate our ideas, but these concepts extend beyond this specific case. For example, we could apply the same approach to determine the optimal placement and number of mobile phone antennas or other amenities and infrastructure units.

\section{Final Remarks}

In the present work, we analyze a simple and instructional case of urban scaling.
It is simple because it considers a city that extends in one-dimensional space only, and it is instructional because this simplicity makes the math more comprehensible and the processes more imaginable.
When we analyze a more realistic city with a (fractal) dimension larger than 1, the calculations for the interaction function become more complicated. However, the fundamental results (sublinear and linear scaling according to $\gamma$ values, and consequently long- and short-range regimes) remain consistent.

Considering human interaction, we essentially show that when accessibility decreases slower than the gain of distance, an agglomeration effect occurs, leading to increasing returns to scale.
In parallel, a scaling economy in infrastructure also occurs, according to which an amenity number less than proportional is sufficient to serve the population. 
This is precisely what complex systems are about -- interacting parts that, in conjunction, are more than individually.

We hope the findings presented here can inspire investigations of real-world cities that approximate a one-dimensional structure. Cities of this type, such as those illustrated in  Fig.~\ref{fig:the_line}, provide an intriguing opportunity to explore how amenities or wealth production could be quantitatively assessed to validate and refine the proposed theoretical framework. If the theory proves to be applicable, it could inform the design and optimization of emerging artificial cities, such as \emph{The Line}. For example, insights from this study could be used to enhance urban efficiency by leveraging transportation infrastructure -- adjusting train speed, frequency, or connectivity to increase the likelihood and number of interactions among individuals.

In conclusion, the results presented in this work provide both theoretical and practical insights into how the spatial organization and interactions among individuals in approximately one-dimensional cities can be optimized. Through the proposed model, we demonstrate that adjustments in the range of interaction (characterized by the decay exponent $\gamma$) can significantly influence connectivity, which results in urban efficiency and wealth production. 
Of course, as mentioned in the introduction, we discuss the approximate one-dimensional cities for illustrative purposes.
The considerations can be generalized to two (or fractal) dimensions, but the one-dimensional situation is analytically simpler and more tangible.
A challenge, however, is the third dimension, which is how to factor in the vertical extent of cities.

\vskip 1 cm

{\bf Acknowledgement:} F.\ L.\ Ribeiro thanks CNPq (grant numbers 403139/2021-0 and 424686/2021-0) and Fapemig (grant number APQ-00829-21) for financial support. 
R.\ L.\ Fagundes thanks Fapemig (grant number APQ-00829-21) for financial support. 
R.\ L.\ Fagundes also appreciates funding from the German Research Foundation (DFG) project UPon (\#451083179) and from the Leibniz project CriticaLand.
D.\ Rybski acknowledges financial support from DFG for the projects UPon (\#451083179) and Gropius (\#511568027).


\newcommand{\etalchar}[1]{$^{#1}$}

\end{document}